\documentclass[11pt]{article}
\usepackage{jheppubmod}
\usepackage{cite}
\pdfoutput=1
\usepackage{cancel}
\usepackage{xcolor}
\usepackage{caption}  
\usepackage{cite}
\usepackage{relsize}
\usepackage{physics}
\usepackage{psfrag}
\usepackage{cancel}
\usepackage{array}
\usepackage{amssymb}
\usepackage{amsmath}
\usepackage[compat=1.1.0]{tikz-feynman}
\usepackage{amsthm}
\usepackage{float}

\usepackage{tikz}
\usetikzlibrary{decorations.markings}

\usepackage{tikz,lipsum,lmodern}
\usepackage[most]{tcolorbox}
\usepackage{hyperref}
\usepackage{xcolor}
\hypersetup{
	colorlinks=true,
	urlcolor=cyan,
	filecolor=green,
	citecolor=green,
	linkcolor=magenta,
}

\pdfoutput=1

\title{Celestial amplitude for 2d theory}
\author[]{Sarthak Duary}
\affiliation[]{International Centre for Theoretical Sciences-TIFR,
Shivakote, Hesaraghatta Hobli, Bengaluru North 560089, India}
\emailAdd{sarthak.duary@icts.res.in}
\date{}

\abstract{We explore celestial amplitude corresponding to $2d$ bulk $\mathcal{S}$-matrix. We consider scalar particles with identical mass and show that the celestial amplitude becomes the fourier transform of the $2d$ $\mathcal{S}$-matrix written in the rapidity variable. We translate the crossing and unitarity conditions into the conditions on the celestial amplitude. For the $2d$ Sinh-Gordon model, we calculate the celestial amplitude perturbatively in coupling constant and check that the crossing and unitarity conditions are satisfied for the celestial amplitude. Imposing the crossing and unitarity conditions to the celestial amplitude, we want to find amplitudes to the higher order in perturbation theory from the lower order i.e., to provide a ``\textit{proof of principle}'' to show we can apply the bootstrap idea to the celestial amplitude. We find that imposing the crossing and unitarity conditions is not enough for bootstrapping celestial amplitude, there is an extra term which can't be fixed by the crossing and unitarity conditions. We also study the gravitational dressing condition in $2d$ QFT for massless particles in celestial space and see that for the gravitationally dressed celestial amplitude, the poles on the right half-plane get erased for several ansatzes.}

\begin{document}
\maketitle
\section{Introduction}
While evaluating Quantum field theory (QFT) scattering amplitudes, we generally express the asymptotic states in terms of asymptotic plane wave solutions to the free wave equation which is energy-momentum eigenstates. This conventional plane wave basis makes spacetime translation invariance manifest due to energy-momentum conservation but obscures conformal invariance. 

The Lorentz group in $\mathbb{R}^{1,d-1}$ is the same as the Euclidean conformal group in $(d-2)$-dimensions. This indicates that the $d$-dimensional $\mathcal{S}$-matrix is related to $(d-2)$-dimensional conformal field theory correlation function. 

\begin{figure}[H]
\centering
\includegraphics[width=0.7\linewidth]{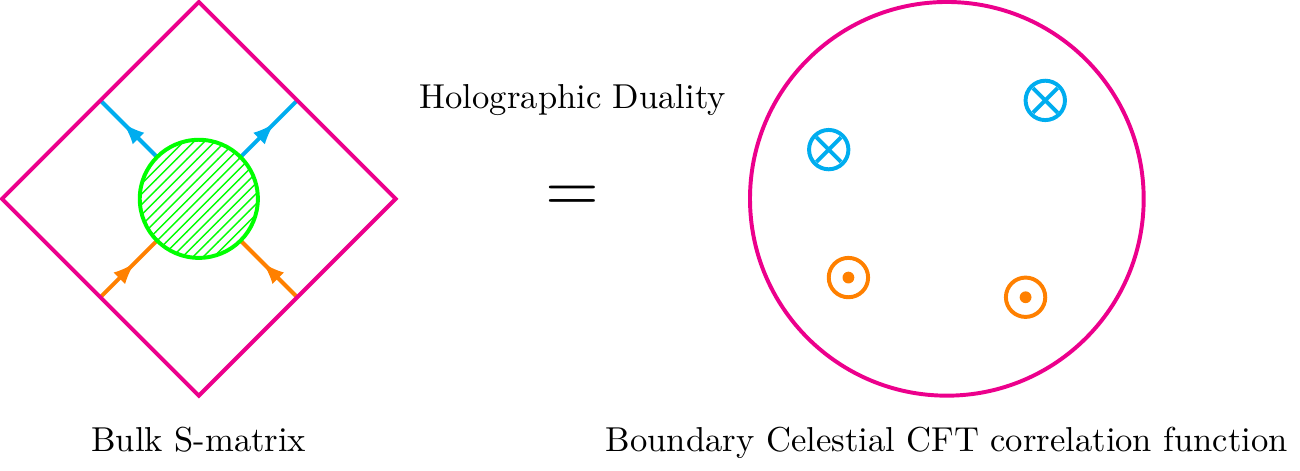}
\caption{Bulk $\mathcal{S}$-matrix mapped to Boundary Celestial CFT correlation function}
\label{fig:img1}
\end{figure}

As opposed to the plane wave basis which is energy-momentum eigenstate, in \cite{Pasterski:2016qvg, Pasterski:2017kqt}, a new basis called the conformal primary basis which is the boost eigenstate is constructed for both massive and massless particles. In this basis, free fields transform as conformal primaries under the Lorentz group and the $\mathcal{S}$-matrix elements transform manifestly as conformal correlation functions on the celestial sphere i.e., the sphere at null infinity. The scattering amplitude computed in this basis are known as celestial amplitude and we can express the bulk $\mathcal{S}$-matrix in terms of a boundary correlation function of `celestial CFT' that lives on the sky as in fig.\ref{fig:img1}.      

The motivation for seeking this conformal basis is to understand the holographic nature of quantum gravity in asymptotically flat spacetimes. Realization of the holographic duality \cite{Maldacena:1997re, Witten:1998qj, Gubser:1998bc} from the bottom up by finding the symmetries that both sides of the holographic dual pair obey \cite{Brown:1986nw, Brown:1992br} gives rise to the enhancement of the Lorentz symmetry to full Virasoro \cite{Kapec:2014opa} and the existence of a stress tensor in a $2d$ CFT obeying Ward identity constructed from subleading soft-graviton mode in the bulk \cite{Kapec:2016jld}. These observations along with the equivalence between soft theorems and asymptotic symmetries i.e., soft theorems recasted as conservation laws associated with large gauge symmetries lead to the proposal that there exists a holographic duality between the theory of gravity in four-dimensional asymptotically flat spacetimes and some sort of exotic CFT living on the two-dimensional celestial sphere at null infinity.

\par 
The flat space holography is initiated by the work of de Bohr and Solodukhin \cite{deBoer:2003vf}. From dS and AdS slicing of Minkowski space, they postulate that the flat space in four dimensions has some description in terms of a theory living on the boundary of these dS and AdS slices which is identified with the celestial sphere.

\par
In this celestial holography paradigm, we study the celestial amplitude for the $2d$ bulk scattering. We construct the map to evaluate celestial amplitude from bulk $\mathcal{S}$-matrix which is implemented by the change of basis from energy-momentum eigenstates to boost eigenstates. To summarize, for massive scalar particles the celestial amplitude is the fourier transform of the $2d$ $\mathcal{S}$-matrix written in the rapidity variable. We calculate the perturbative celestial amplitude for the $2d$ Sinh-Gordon model. One subtle point here is that in $2d$, the perturbative expansion of the Sinh-Gordon $\mathcal{S}$-matrix contains pole at rapidity $\theta=0$. Therefore while evaluating the perturbative celestial amplitude, we should implement $i\epsilon$ prescription and as a consequence of it there are two types of celestial amplitude which we indicate by the retarded and advanced celestial amplitude. In the celestial space, we map the crossing and unitarity conditions to calculate the celestial amplitude to the higher order in perturbation theory from the lower one by imposing these constraints which we refer as ``celestial bootstrap''. We check the crossing and unitarity conditions for the Sinh-Gordon model.

\par 
In AdS, solving conformal crossing equation, the one-loop correction to the four point amplitude is evaluated for scalar $\phi^4$ theory in \cite{Aharony:2016dwx} from which one can analytically calculate the anomalous dimensions. To apply the CFT-bootstrap technique in flat space to find the amplitude, we map the crossing and unitarity conditions in the celestial space for $2d$ scattering. For massive scalar case, in higher than $2d$, technically it's a bit challenging to evaluate the celestial amplitude even at tree level say for massive $\phi^4$ theory in $4d$, for that as a stepping stone towards bootstrapping celestial amplitude we restrict to $2d$ scattering.
 
\par 
We also find that imposing the crossing and unitarity conditions to the celestial amplitude is not enough to determine the higher-order perturbative celestial amplitude from the lower order. There is an extra term which can't be fixed by the crossing and unitarity conditions. 


\par 
We study the gravitational dressing condition for $2d$ QFT amplitude for massless particles in celestial space. We see that after the gravitational dressing in celestial space, the poles on the right half-plane of the celestial amplitude get erased for the functions having multiple poles on the right half-plane.  

\par 
In the conclusion, authors of \cite{Garcia-Sepulveda:2022lga} commented on studying the celestial amplitude for the bulk $2d$ spacetime, which we initiate in this work.

\par 
\subsection*{Outline:} The paper is organized as follows. In section \ref{mcamp}, we review the construction of massive celestial amplitude. In section \ref{ctwo}, we define the celestial $4$-point amplitude for $2 \to 2$ scattering of massive scalar particles in $2d$. The massive celestial amplitude in $2d$ is given by the fourier transform of the $\mathcal{S}$-matrix written in the rapidity variable. In section \ref{Sinh}, we evaluate the celestial amplitude in $2d$ Sinh-Gordon model by perturbatively expanding the $\mathcal{S}$-matrix. In section \ref{5}, we translate the crossing and unitarity conditions in celestial space and check the crossing and unitarity conditions for the Sinh-Gordon model. In section \ref{6}, we discuss about reconstructing the higher-order correction to the $4$-point celestial amplitude from the lower-order amplitude using the ``crossing and unitarity'' conditions. In section \ref{gravdres}, we translate the gravitational dressing condition of the $2d$ QFT amplitude in celestial space. For massless particles, we see that the gravitational dressing condition translated to celestial space acts as a pole eraser from the right half-plane corresponding to the celestial amplitude for the functions having multiple poles on the right half-plane. In section \ref{conc}, we summarize our main results and discuss open problems and future directions.

\newpage 
\section{Massive Celestial amplitude}
\label{mcamp}
In this section, we review the construction of massive celestial amplitude described in \cite{Pasterski:2017kqt}.

The massive scalar conformal primary wavefunction $\phi_{\Delta}^{\pm}(X^{\mu};\vec{w})$ in $\mathbb{R}^{1,d-1}$ admits the fourier expansion on the plane waves 

\begin{equation}
\label{map}
\phi_{\Delta}^{\pm}(X^{\mu};\vec{w})=\int_{H_{d-1}} [d\hat{p}] ~~G_{\Delta}(\hat{p};\vec{w})~\text{exp}[\pm ~im~\hat{p}.X]~~~,
\end{equation}

where the on-shell momenta per mass, a unit timelike vector $\hat{p}(y,\vec{z})$ satisfying $\hat{p}^2=-1$ can be parametrized using the $H_{d-1}$ coordinates $y$ ($y>0$) and $z \in \mathbb{R}^{d-2}$ as,

\begin{equation}
\hat{p}(y,\vec{z})=\Bigg(\frac{1+y^2+|\vec{z}|^2}{2y},\frac{\vec{z}}{y},\frac{1-y^2-|\vec{z}|^2}{2y}\Bigg)~~~.
\end{equation} 
  
Here, $[d \hat{p}]$  is the $SO(1, d-1)$ invariant measure on $H_{d-1}$
\begin{equation}
\begin{split}
\int_{H_{d-1}}^{} [d\hat{p}] &\equiv \int \frac{d^{d-1}\hat{p}^i}{\hat{p}^0}\\
&= \int_{0}^{\infty}\frac{dy}{y^{d-1}} \int d^{d-2}\vec{z}~~~,~~~i=1, \ldots, d-1 ~~~, ~~~\hat{p}^0=\sqrt{\hat{p}^i\hat{p}^i+1}~~~.
\end{split}
\end{equation}
  
Here, $G_{\Delta}(\hat{p};\vec{w})$ is the scalar bulk-to-boundary propagator in $H_{d-1}$ given by \cite{Witten:1998qj}
\begin{equation}
G_{\Delta}(\hat{p};\vec{w})=\Bigg(\frac{y}{y^2+|\vec{z}-\vec{w}|^2}\Bigg)^{\Delta}~~~,
\end{equation}
where $\vec{w} \in \mathbb{R}^{d-2}$ lies on the boundary of $H_{d-1}$. The scalar bulk-to-boundary propagator written in terms of $\hat{p}^{\mu}(y,\vec{z})$ and null momentum $q^{\mu}(\vec{w})=(1+|\vec{w}|^2, 2\vec{w}, 1-|\vec{w}|^2)$ in $\mathbb{R}^{1,d-1}$ is given by \cite{Costa:2014kfa} 

\begin{equation}
G_{\Delta}(\hat{p};q)=\frac{1}{(-\hat{p}.q)^{\Delta}}~~~.
\end{equation}

Using the mapping from the plane wave to the conformal primary wavefunction given by eq.\eqref{map}, the $\mathcal{S}$-matrix elements in the conformal prinary basis is given in terms of an integral transform

\begin{equation}
\label{cel}
\mathcal{\tilde{A}}(\Delta_i,\vec{w}_i)=\prod_{k=1}^{n}\int_{H_{d-1}}^{} [d\hat{p}_k] G_{\Delta_k}(\hat{p}_k;\vec{w}_k) \mathcal{A}(\pm m_i \hat{p}^{\mu}_i)~~~,
\end{equation}

where $\pm m_i \hat{p}^{\mu}_i$ parametrization depends on whether the particle is incoming or outgoing.

In the r.h.s., $\mathcal{A}(\pm m_i \hat{p}^{\mu}_i)$ is the momentum space amplitude along with the momentum conserving delta function.

We define $\mathcal{\tilde{A}}(\Delta_i,\vec{w}_i)$ by the massive celestial amplitude. Under the conformal symmetry action the massive celestial amplitude transforms covariantly as a $(d-2)$-dimensional CFT $n$-point function of scalar primaries with dimension $\Delta_i$
\begin{equation}
\mathcal{\tilde{A}}(\Delta_i,\vec{w}^{\prime}_i(\vec{w}_i))=\prod_{k=1}^{n}\Bigg| \frac{\partial \vec{w}^{\prime}_k}{\partial \vec{w}_k}\Bigg|^{\frac{-\Delta_k}{d-2}}\mathcal{\tilde{A}}(\Delta_i, \vec{w}_i)~~~.
\end{equation}  
  
\parskip=10pt
\section{Massive Celestial amplitude for ~$2 \to 2$ scattering: The Celestial point}
\label{ctwo}
In this section, we define the celestial $4$-point amplitude for $2 \to 2$ scattering of massive scalar particles in $2d$.

In $2d$, the on-shell momenta of particles of mass $m$ can be written in the rapidity parametrization $\theta$
\begin{equation}
\begin{split}
p^{\mu}&=(p^{0}=E,~ p^{1}=p)\\
&=m(\cosh \theta, \sinh\theta)~~~.
\end{split}
\end{equation}

We consider the elastic scattering process of identical real scalar particles of mass $m$ with rapidities $\theta_1$ and $\theta_2$. 


Energy and momentum conservation for identical particles give
\begin{equation}
\begin{split}
&\theta_1=\theta_4~~,~~\theta_2=\theta_3~~~.\\
\end{split}
\end{equation}
The $2 \to 2$ $\mathcal{S}$-matrix is given by
\begin{equation}
\label{amp1}
\begin{split}
\mathcal{S}_{2 \to 2}=\langle \theta_4,\theta_3 \ket{\theta_1,\theta_2}&=4p_1^0 p_2^0 \times (2 \pi)^2 \delta(p_1^1-p_4^1)\delta(p_2^1-p_3^1)+i (2\pi)^2 \delta^{(2)}(p_1^{\mu}+p_2^{\mu}-p_3^{\mu}-p_4^{\mu})~\mathcal{T}\\
&=4 (2\pi)^2\delta(\theta_1-\theta_4) \delta(\theta_2-\theta_3) \Bigg(1+\frac{i~\mathcal{T}(\theta_1-\theta_2)\csch(\theta_1-\theta_2)}{4m^2}\Bigg)~~~.
\end{split}
\end{equation}

We define $\mathcal{S}$-matrix which is dependent on the difference of rapidities as 
\begin{equation}
S(\theta_1-\theta_2) \equiv 1+\frac{i \mathcal{T}(\theta_1-\theta_2) \csch (\theta_1-\theta_2)}{4m^2}~~~.\nonumber
\end{equation}

The Mandelstem variables are

\begin{equation}
s=(p_1+p_2)^2=4m^2 \cosh^2\Big(\frac{\theta}{2}\Big)~~~,~~~t=4m^2-s~~~,~~~u=0~~~.\nonumber 
\end{equation}
$s$ is the cenrer of mass energy squared, $t$-channel gives $\theta \to i\pi-\theta$.


In $2d$, the celestial $4$-point amplitude of the massive conformal primary wavefunction is 
\begin{equation}
\label{celamp}
\tilde{\mathcal{A}}=\Bigg(\prod_{i=1}^{4}\int_{} \frac{d\hat{p}_i^1}{\hat{p}_i^0}\Bigg) \times \prod_{i=1}^{4}G_{\Delta_i}(\hat{p}_i)~ \mathcal{S}_{2 \to 2}
\end{equation}
where, $\hat{p}_i^{\mu} \equiv \hat{p}^{\mu}(\theta_i) = \frac{p_i^{\mu}}{m}=(\cosh \theta_i, \sinh\theta_i)$ and $G_{\Delta_i}(\hat{p}_i)$ is the bulk-to-boundary propagator in $H_1$.


The scalar bulk-to-boundary propagator in $H_1$ is
\begin{equation}
\begin{split}
G_{\Delta}(\hat{p};q)&=\frac{1}{(-\hat{p}.q)^{\Delta}}=\frac{1}{(\cosh\theta-\sinh\theta)^{\Delta}}\\
&=e^{\Delta \theta}~~~.
\end{split}
\end{equation}   
Here, $q=(1,1)$
since, conformal boundary of $H_1$ are specified by $2$ dimesnional points on the projective null cone
$$-(q^0)^2+(q^1)^2=0~~,~~~q \sim \lambda q~~~.$$ 

The massive scalar conformal primary wavefunctions in $d$ dimensions are delta-function normalizable when $\Delta$ belongs to the principal continuous series of the irreducible unitary $SO(1,d-1)$ representations, 

$$\Delta \in \frac{d-2}{2} +i \mathbb{R}~~.$$    
Therefore, for the $2d$ scattering, $\Delta$ is pure imaginary. Now, we redefine $i \Delta$ as the conformal dimension of the conformal primary wavefunction in $2d$. The $H_1$ scalar bulk-to-boundary propagator $G_{\Delta}(\hat{p})=e^{i\Delta \theta}$. 

The normalization condition for the $H_1$ scalar 
bulk-to-boundary propagator is 

\begin{equation}
\int_{-\infty}^{\infty} d\theta e^{i \Delta_i \theta}  e^{i \Delta_j \theta} =2\pi ~\delta( \Delta_i+\Delta_j) ~~~.
\end{equation}




Now, the integral measure is 
\begin{equation}
\prod_{i=1}^{4}\int \frac{d\hat{p}_i^1}{\hat{p}_i^0}=\prod_{i=1}^{4} \int_{-\infty}^{\infty} d\theta_i~~~.\nonumber  
\end{equation}

We can express celestial $4$-point amplitude in eq.\eqref{celamp} as 
\begin{equation}
\small 
\begin{split}
\tilde{\mathcal{A}}
&=4(2\pi)^2 \prod_{i=1}^{4} \int_{-\infty}^{\infty} d\theta_i 
~~e^{i\Delta_i \theta_i}S(\theta_1-\theta_2)\delta(\theta_1-\theta_4) \delta(\theta_2-\theta_3)\\
&=4(2\pi)^2 \int_{-\infty}^{\infty} \int_{-\infty}^{\infty} d\theta_4~ d\theta_3~S(\theta_4-\theta_3)~e^{i\Delta_1 \theta_4}e^{i\Delta_2 \theta_3}e^{i\Delta_3 \theta_3}e^{i\Delta_4 \theta_4}\\
&=4(2\pi)^2\frac{1}{2}\int_{-\infty}^{\infty} \int_{-\infty}^{\infty} d\theta_+~ d\theta_-~S(\theta_-)e^{\frac{i}{2}(\Delta_1+\Delta_4+\Delta_2+\Delta_3)\theta_+}e^{\frac{i}{2}(\Delta_1+\Delta_4-\Delta_2-\Delta_3)\theta_-}\\
&=16 \pi^3 ~\delta(\Delta_+)~\mathcal{A}(\Delta_-)~~~.
\end{split}
\end{equation}
where, we define light-cone coordinates $\theta_{\pm}=\theta_4\pm \theta_3$, $\Delta_{\pm}=\frac{1}{2}(\Delta_1+\Delta_4\pm \Delta_2\pm \Delta_3)$ and $$\mathcal{A}(\Delta_-)\equiv \int_{-\infty}^{\infty} d\theta e^{i \Delta_- \theta_-} S(\theta_-)~~~.$$

Now, we strip off the overall delta function $\delta(\Delta_+)$ and  
name the conjugate pair of $\theta_-$ as $\omega\equiv \Delta_-=\frac{1}{2}(\Delta_1+\Delta_4-\Delta_2-\Delta_3)~,$ the $2d$ celestial amplitude is the fourier transform of the $S$-matrix with respect to rapidity.
\begin{equation}
\label{fourt}
\mathcal{A}(\omega)\equiv \int_{-\infty}^{\infty} d\theta e^{i \omega \theta} S(\theta)~~~.
\end{equation}

Physically, the rapidity $\theta$ shifts under the boost as $\theta\to \theta+c$. Therefore, in order to diagonalize the boost which is achieved by the conformal basis, we need to perform the fourier transform of the $\mathcal{S}$-matrix $S(\theta)$ with respect to rapidity. 

Here, there is no celestial coordinate and the dual theory is zero-dimensional which we refer as the ``celestial point''. The CFT correlation function ``dual'' to the bulk $\mathcal{S}$-matrix would be a zero-dimensional CFT correlation function or said differently, an operator algebra without coordinates.

	
	
	
%

\section{Celestial amplitude in 2d Sinh-Gordon model}
\label{Sinh}
In this section, we evaluate the celestial amplitude in $2d$ Sinh-Gordon model by perturbatively expanding the $\mathcal{S}$-matrix of the Sinh-Gordon model.

The largangian of the Sinh-Gordon model \cite{Rosenhaus:2019utc,Arefeva:1974bk} is
\begin{equation}
\mathcal{L}=\frac{1}{2}(\partial \phi)^2+\frac{m^2}{b^2}(\cosh (b\phi)-1)~~~.
\end{equation}
The $\mathcal{S}$-matrix for Sinh-Gordon model with the parameter $\alpha$ related to the coupling $b$ is 
\begin{equation}
S(\theta)=\frac{\sinh\theta-i \sin\alpha}{\sinh \theta+i \sin\alpha}~~~,~~~\alpha=\frac{\pi b^2}{8 \pi +b^2}~~~.
\end{equation}
We can verify the $\mathcal{S}$-matrix by expanding the lagrangian perturbatively in coupling $b$
\begin{equation}
\mathcal{L}=\frac{1}{2}(\partial \phi)^2+\frac{1}{2} m^2\phi^2+\frac{m^2b^2}{4!}\phi^4+\frac{m^2b^4}{6!}\phi^6+\cdots~~~,
\end{equation}
and calculate the perturbative expansion of the $\mathcal{S}$-matrix to first few orders in $b^2$. 

The perturbative expansion of the $\mathcal{S}$-matrix with respect to $b^2$ is given by 
\begin{equation}
\label{sinhg}
\small 
\begin{split}
S^{(0)}(\theta)&=1\\
S^{(1)}(\theta)&=-\frac{1}{4} i b^2 \text{csch}\theta \\
S^{(2)}(\theta)&=-\frac{b^4 \text{csch}\theta  (\pi  \text{csch}\theta -i)}{32 \pi }\\
S^{(3)}(\theta)&=\frac{i b^6 \text{csch}\theta  \left(6 \pi ^2 \text{csch}^2\theta -12 i \pi  \text{csch}\theta +\pi ^2-6\right)}{1536 \pi ^2}\\
S^{(4)}(\theta)&=\frac{b^8 \text{csch}\theta  \left(6 \pi ^3 \text{csch}^3\theta -18 i \pi ^2 \text{csch}^2\theta +2 \pi  \left(\pi ^2-9\right) \text{csch}\theta -3 i \left(\pi ^2-2\right)\right)}{12288 \pi ^3}\\
S^{(5)}(\theta)&=-\frac{ib^{10} \csch\theta\left(120\pi^4\text{csch}^4\theta-480i\pi^3\text{csch}^3\theta+60\pi^2(\pi^2-12)\text{csch}^2\theta-160i\pi(\pi^2-3)\text{csch}\theta+\pi^4-120\pi^2+120\right)}{1966080 \pi^4}.
\end{split}
\end{equation}

Since, the perturbative expansion of the $\mathcal{S}$-matrix $S(\theta)$ contains poles at $\theta=0$, we define the celestial amplitude using $i\epsilon$ prescription as $$\mathcal{A}^{\pm}(\omega)\equiv \int_{-\infty}^{\infty} d\theta e^{i \omega \theta} S(\theta\pm i\epsilon)$$ 
where 
\begin{equation}
\label{pertcel}
\mathcal{A}^{\pm}(\omega)=2\pi \Big[\delta(\omega)+b^2f_1^{\pm}(\omega)+b^4f_2^{\pm}(\omega)+\cdots\Big]~~~, 
\end{equation}



\begin{equation}
\begin{split}
f_n^{\pm}(\omega)=\frac{1}{2\pi (b^2)^n}\int_{-\infty}^{\infty} d\theta e^{i \omega \theta} S^{(n)}(\theta\pm i\epsilon)~~~.
\end{split}
\end{equation}

Here, we call $\mathcal{A}^{+}(\omega)$ as the retarded celestial amplitude and $\mathcal{A}^{-}(\omega)$ as the advanced celestial amplitude. 

Here, eq.\eqref{pertcel} can be understood as the perturbative expansion of the retarded and advanced celestial amplitude and we call $f_n^{+}(\omega)$ and $f_n^{-}(\omega)$ as the perturbative retarded and advanced celestial amplitude. 

Now, we evaluate the fourier transform by $+i \epsilon$ prescription. We shift the pole at $\theta=0$ to $\theta=-i\epsilon$ and enclose the contour in the upper half-plane.

Evaluating the fourier transform using $+i\epsilon$ prescription, we get the perturbative retarded celestial amplitude   
\begin{equation}
\small 
\begin{split}
f_1^+(\omega)&=-\frac{ 1 }{4(1+e^{\pi \omega})}\\
f_2^+(\omega)&=\frac{\left(\pi  \omega +e^{\pi  \omega } (\pi  \omega +1)-1\right) (\coth (\pi  \omega )-1)}{64 \pi }\\
f_3^+(\omega)&=-\frac{ e^{\frac{\pi  \omega }{2}} (\coth (\pi  \omega )-1) \left[\left(\pi ^2 \left(6 \omega ^2+5\right)+6\right) \sinh \left(\frac{\pi  \omega }{2}\right)+12 \pi  \omega  \cosh \left(\frac{\pi  \omega }{2}\right)\right]}{1536 \pi^2}\\
f_4^+(\omega)&=\frac{ e^{\frac{\pi  \omega }{2}} (\coth (\pi  \omega )-1) \left[3 \left(\pi ^2 \left(6 \omega ^2+5\right)+2\right) \sinh \left(\frac{\pi  \omega }{2}\right)+\pi  \omega  \left(\pi ^2 \left(\omega ^2+2\right)+18\right) \cosh \left(\frac{\pi  \omega }{2}\right)\right]}{12288 \pi^3}\\
f_5^+(\omega)&=-\frac{ e^{\frac{\pi  \omega }{2}} (\coth (\pi  \omega )-1)}{1966080  \pi^4}\times \\
& \Bigg[\left(120 \pi ^2 \left(6 \omega ^2+5\right)+\pi ^4 \left(5 \omega ^2 \left(\omega ^2-2\right)-14\right)+120\right) \sinh \left(\frac{\pi  \omega }{2}\right)+80 \pi  \omega  \left(\pi ^2 \left(\omega ^2+2\right)+6\right) \cosh \left(\frac{\pi  \omega }{2}\right)\Bigg]~~~.
\end{split}
\end{equation}

Next, we evaluate the integral by $-i \epsilon$ prescription. We shift the pole at $\theta=0$ to $\theta=i\epsilon$ and enclose the contour in the lower half-plane.

Evaluating the fourier transform using $-i\epsilon$ prescription, we get the perturbative advanced celestial amplitude 
\begin{equation}
\small 
\begin{split}
f_1^-(\omega)&=\frac{1}{4 (e^{-\pi  \omega }+1)}\\
f_2^-(\omega)&=\frac{e^{\pi  \omega } \left(\pi  \omega +e^{\pi  \omega } (\pi  \omega -1)+1\right) (\coth (\pi  \omega )-1)}{64 \pi }\\
f_3^-(\omega)&=\frac{e^{\pi  \omega } (\coth (\pi  \omega )-1) \left[e^{\pi  \omega } \left(\pi ^2 \left(3 \omega ^2+2\right)-12 \pi  \omega +6\right)-3 \pi  \omega  (\pi  \omega +4)-2 \left(3+\pi ^2\right)\right] }{3072 \pi ^2}\\
f_4^-(\omega)&=\frac{e^{\pi  \omega }  (\coth (\pi  \omega )-1) \left[\pi  (\pi  \omega +3) \left(\pi  \left(\omega ^2+2\right)+6 \omega \right)+e^{\pi  \omega } \left(\pi  (\pi  \omega -3) \left(\pi  \left(\omega ^2+2\right)-6 \omega \right)-6\right)+6\right]}{24576 \pi ^3}\\
f_5^-(\omega)&=\frac{e^{\frac{3 \pi  \omega }{2}} (\coth (\pi  \omega )-1)}{1966080 \pi ^4} \times 
\\&\Bigg[\left(120 \pi ^2 \left(3 \omega ^2+2\right)+\pi ^4 \left(5 \left(\omega ^2+4\right) \omega ^2+16\right)+120\right) \sinh \left(\frac{\pi  \omega }{2}\right)-80 \pi  \omega  \left(\pi ^2 \left(\omega ^2+2\right)+6\right) \cosh \left(\frac{\pi  \omega }{2}\right)\Bigg].
\end{split}
\end{equation}
In appendix \ref{pert}, we give the details of the computation of the perturbative retarded and advanced celestial amplitude.

\newpage 
\section{Crossing and unitarity conditions in celestial space}
\label{5}
In this section, our aim is to translate the crossing and unitarity conditions into the conditions on the celestial amplitude.

In terms of rapidity, $\theta$ we express the crossing and the unitarity conditions as \cite{Paulos:2016but, Rosenhaus:2019utc, Dorey:1996gd, Pedro}

\begin{equation}
\begin{split}
&S(\theta)=S(i \pi -\theta)\\
&|S(\theta)|^2 = 1~~~,
\end{split}
\end{equation}
where $\theta=\theta_1-\theta_2$~.

The crossing condition physically implies the symmetry of the $\mathcal{S}$-matrix under the exchange of the $s$ and $t$ channels. Unitarity condition physically implies the probablity of getting $2$-particle final state given initial state should be less than or equal to one, i.e., $|S(\theta)|^2 \leq 1$. Assuming integrability condition,  we have $|S(\theta)|^2 = 1.$




The crossing condition in celestial space is obtained by taking the fourier transform of both sides of the crossing condition in the rapidity variable. 
Since, the perturbative expansion of the $\mathcal{S}$-matrix $S(\theta)$ contains poles at $\theta=0$, therefore, perturbatively if we expand upto a given order we should take the fourier transform of $S(\theta \pm i\epsilon)$. 

The crossing condition in celestial space is obtained by taking the fourier transform of both sides 
\begin{equation}
\begin{split}
\int_{-\infty}^{\infty} d\theta e^{i \omega \theta} S(\theta+i\epsilon)&=\int_{-\infty}^{\infty} d\theta e^{i \omega \theta} S(i\pi-\theta+i\epsilon)\\
&=-\int_{i\pi+\infty}^{i\pi-\infty} d\theta^{\prime} e^{i \omega (i\pi-\theta^{\prime})} S(\theta^{\prime}+i\epsilon)~~~( i\pi-\theta\equiv \theta^{\prime})\\
&=\int_{i\pi-\infty}^{i\pi+\infty} d\theta^{\prime} e^{i \omega (i\pi-\theta^{\prime})} S(\theta^{\prime}+i\epsilon)\\
&=e^{-\omega \pi} \int_{-\infty}^{+\infty} d\theta^{\prime} e^{-i \omega\theta^{\prime}} S(\theta^{\prime}+i\epsilon)\\
&=e^{-\omega \pi}\mathcal{A}^{+}(-\omega)\\
\implies \mathcal{A}^{+}(\omega)&=e^{-\omega \pi}\mathcal{A}^{+}(-\omega)~~~.
\end{split}
\end{equation}

where, in the last step we use 
\begin{equation}
\int_{i\pi-\infty}^{i\pi+\infty} d\theta^{\prime} e^{i \omega (i\pi-\theta^{\prime})} S(\theta^{\prime})=\int_{-\infty}^{+\infty} d\theta^{\prime} e^{i \omega (i\pi-\theta^{\prime})} S(\theta^{\prime})~~~.
\end{equation}
which is valid when we have no poles in the physical strip $0<\text{Im}(\theta)<\pi.$ 


Similarly, the crossing condition while taking the fourier transform of $$\int_{-\infty}^{\infty} d\theta e^{i \omega \theta} S(\theta-i\epsilon)$$ becomes 
\begin{equation}
\begin{split}
\int_{-\infty}^{\infty} d\theta e^{i \omega \theta} S(\theta-i\epsilon)&=\int_{-\infty}^{\infty} d\theta e^{i \omega \theta} S(-i\pi-\theta-i\epsilon)\\
\implies \mathcal{A}^{-}(\omega)&=e^{\omega \pi}\mathcal{A}^{-}(-\omega)~~~.
\end{split}
\end{equation} 

The crossing condition relates the retarded (advanced) celestial amplitude of positive $\omega$ to the retarded (advanced) celestial amplitude of negative $\omega$ and vice-versa.




Perturbatively expanding $\mathcal{A}^{\pm}(\omega)$ 
as
\begin{equation}
\mathcal{A}^{\pm}(\omega)=2\pi \Big[\delta(\omega)+b^2f_1^{\pm}(\omega)+b^4f_2^{\pm}(\omega)+b^6f_2^{\pm}(\omega)+\cdots\Big]~~~,\nonumber 
\end{equation}
$f_n^{\pm}(\omega)$ satisfy the crossing condition 
\begin{equation}
f_n^{\pm}(\omega)=e^{\mp \omega \pi}f_n^{\pm}(-\omega)~~~.
\end{equation}

The unitarity condition gives 

\begin{equation}
\begin{split}
&S(\theta+i \epsilon)S(\theta+i \epsilon)^{*}=1\\
\implies &S(\theta+i \epsilon)S(-\theta-i \epsilon)=1~~~.
\end{split}
\end{equation}

Here, the unitarity condition $|S(\theta)|^2 = 1$ becomes $S(\theta+i \epsilon)S(-\theta-i\epsilon)$. We combine the original unitarity condition with the real analyticity of the $S$-matrix  $$S(\theta+i\epsilon)^{*}=S(-\theta-i\epsilon)~~~.$$.

The unitarity condition in celestial space is obtained by taking the fourier transform of both sides of
$$S(\theta+i \epsilon)S(-\theta-i \epsilon)=1~~~.$$  
Now, the multiplication of functions $S(\theta+i\epsilon) S(-\theta-i \epsilon)$ gets converted into the convolution under the fourier transform as follows
\begin{equation}
\begin{split}
\int_{-\infty}^{\infty} d\theta e^{i \omega \theta}S(\theta+i\epsilon)S(-\theta-i\epsilon)&=\int_{-\infty}^{\infty} d\theta e^{i \omega \theta} S(\theta+i\epsilon) \Bigg[\frac{1}{2\pi} \int_{-\infty}^{\infty} d\omega^{\prime}e^{i\omega^{\prime}\theta}\mathcal{A}^-(\omega^{\prime})\Bigg]\\
&=\frac{1}{2\pi} \int_{-\infty}^{\infty} d\omega^{\prime} \mathcal{A}^{-}(\omega^{\prime})\int_{-\infty}^{\infty} S(\theta+i \epsilon) e^{i(\omega+\omega^{\prime})\theta} d\theta \\
&=\frac{1}{2\pi} \int_{-\infty}^{\infty} d\omega^{\prime} \mathcal{A}^-(\omega^{\prime})\mathcal{A}^+(\omega+\omega^{\prime})~~~.
\end{split}
\end{equation}

Here, $$\mathcal{A}^{\pm}(\omega)\equiv \int_{-\infty}^{\infty} d\theta e^{i \omega \theta} S(\theta\pm i\epsilon)~~~.$$ 

Here, $\mathcal{A}^{+}(\omega)$ is the retarded celestial amplitude which is the fourier transform using $+i\epsilon$ prescription enclosing the contour in the upper half-plane in counterclockwise way and $\mathcal{A}^{-}(\omega)$ is the advanced celestial amplitude which is the fourier transform using $-i\epsilon$ prescription enclosing the contour in the lower half-plane in clockwise way.  

Therefore, the unitarity condition in celestial space becomes
\begin{equation}
\label{unitarity}
\frac{1}{2\pi} \int_{-\infty}^{\infty} d\omega^{\prime} \mathcal{A}^+(\omega+\omega^{\prime})\mathcal{A}^-(\omega^{\prime})=2\pi \delta(\omega)~~~.
\end{equation}
Perturbatively expanding $\mathcal{A}^{\pm}(\omega)$ 
as
\begin{equation}
\mathcal{A}^{\pm}(\omega)=2\pi \Big[\delta(\omega)+b^2f_1^{\pm}(\omega)+b^4f_2^{\pm}(\omega)+b^6f_2^{\pm}(\omega)+\cdots\Big]~~~,\nonumber 
\end{equation}

\begin{equation}
\begin{split}
f_n^{\pm}(\omega)=\frac{1}{2\pi (b^2)^n}\int_{-\infty}^{\infty} d\theta e^{i \omega \theta} S^{(n)}(\theta\pm i\epsilon)~~~,
\end{split}
\end{equation}

and then put it in eq.\eqref{unitarity} we have
the unitarity condition order by order in perturbation theory
\begin{equation}
\begin{split}
f_1^{+}(\omega)+f_1^{-}(-\omega)=0 \nonumber ~~~,
\end{split}
\end{equation}
\begin{equation}
\begin{split}
f_n^{+}(\omega)+f_n^{-}(-\omega)+\sum_{j=1}^{n-1} \int_{-\infty}^{\infty}d\omega^{\prime}   f_{n-j}^{+}(\omega+\omega^{\prime}) f_j^{-}(\omega^{\prime}) =0~~~(n>1)~~~.
\end{split}
\end{equation}

One important thing to note is that while translating the crossing and unitarity conditions in celestial space the retarded and the advanced celestial amplitudes naturally appears. 

Now, since 
\begin{equation}
S^{(n)}(\theta+i\epsilon)+S^{(n)}(-\theta-i\epsilon)=2~\text{Re}~S^{(n)}(\theta+i\epsilon) \nonumber 
\end{equation}
$f_n^{+}(\omega)+f_n^{-}(-\omega)$ is related ro the fourier transform of the real part of $S^{(n)}(\theta+ i\epsilon)$ 
\begin{equation}
f_n^{+}(\omega)+f_n^{-}(-\omega)=2~\frac{1}{2\pi (b^2)^n}\int_{-\infty}^{\infty} d\theta e^{i \omega \theta} \text{Re}S^{(n)}(\theta+ i\epsilon)~~~.
\end{equation}

\begin{figure}[H]
	\centering
	\includegraphics[scale=0.9]{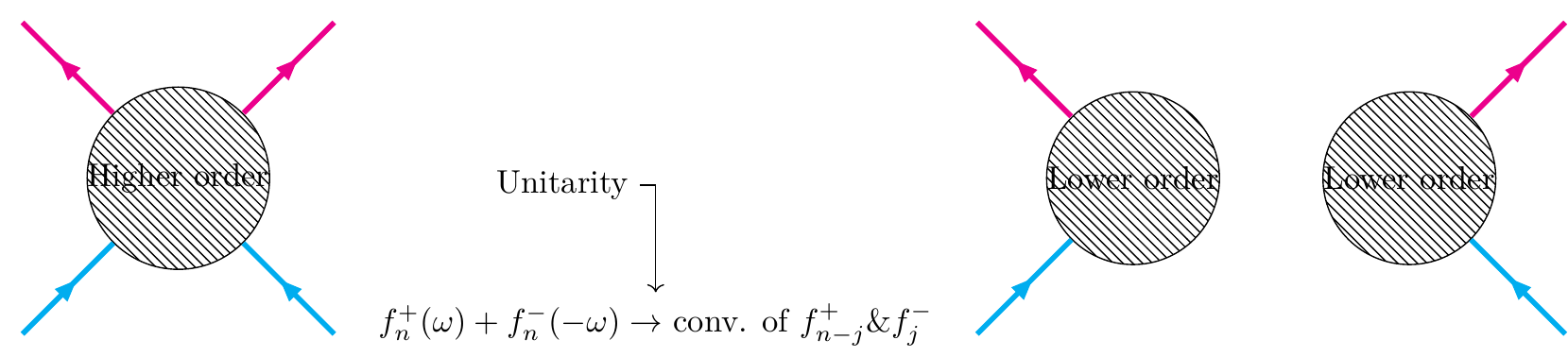}
\caption{Pictorial representation of the unitarity condition in celestial space}
\label{figuni}
\end{figure}

Therefore, we can say that the unitarity condition in perturbation theory translated into celestial amplitude relates the fourier transform of the real part of the perturbative $\mathcal{S}$-matrix at a given order to the convolution of the retarded and advanced celestial amplitude of lower-orders. We illustrate the unitarity condition by the diagram \ref{figuni}.

\subsection{Checking the crossing $\&$ unitarity conditions in celestial space for the 2d Sinh-Gordon model}

We check the crossing and unitarity conditions for the $2d$ Sinh-model using celestial amplitudes obtained in section \ref{Sinh}.

$f_n^{\pm}(\omega)$ satisfy the crossing condition 
\begin{equation}
f_n^{\pm}(\omega)=e^{\mp \omega \pi}f_n^{\pm}(-\omega)~~~.
\end{equation}

The unitarity conditions are satisfied  
\begin{equation}
\begin{split}
f_1^{+}(\omega)+f_1^{-}(-\omega)=0 \nonumber ~~~,
\end{split}
\end{equation}
\begin{equation}
\begin{split}
f_2^{+}(\omega)+f_2^{-}(-\omega)+ \int_{-\infty}^{\infty}d\omega^{\prime}   f_{1}^{+}(\omega+\omega^{\prime}) f_1^{-}(\omega^{\prime}) =0~~~,
\end{split}
\end{equation}
where the integral is given by
\begin{equation}
\int_{-\infty}^{\infty}d\omega^{\prime}   f_{1}^{+}(\omega+\omega^{\prime}) f_1^{-}(\omega^{\prime})=\frac{\omega}{16(1-e^{\pi \omega})}~~~.
\end{equation}

In deriving the crossing and unitarity conditions in celestial space we should be extremely careful about the $i \epsilon$ prescription. 

The convolution of the perturbative retarded and retarded celestial amplitude and the perturbative advanced and advanced celestial amplitude diverges 

$$\int_{-\infty}^{\infty}d\omega^{\prime}   f_{1}^{\pm}(\omega+\omega^{\prime}) f_1^{\pm }(\omega^{\prime}) \rightarrow \infty~~~.$$

The proper $i\epsilon$ prespription involving the convolution of the retarded and advanced celestial amplitude cures this divergence. 

\newpage 
\section{Bootstrapping Celestial amplitude}
\label{6}
In this section, we see that using the crossing and unitarity conditions, how much we can get for the higher order celestial amplitude from the lower order celestial amplitude.

Crossing condition translated in celestial space gives 
\begin{equation}
\label{boot1}
f_n^{\pm}(\omega)=e^{\mp \pi \omega} f_n^{\pm}(-\omega)~~~.
\end{equation}
Unitarity condition translated in celestial space gives 
\begin{equation}
\label{boot2}
f_n^{+}(\omega)+f_n^{-}(-\omega)+\sum_{j=1}^{n-1} \int_{-\infty}^{\infty}d\omega^{\prime}   f_{n-j}^{+}(\omega+\omega^{\prime}) f_j^{-}(\omega^{\prime}) =0~~~.
\end{equation}

Now, 
the relation between $f_n^{+}(\omega)$ and $f_n^{-}(\omega)$ is 
\begin{equation}
\label{diff}
f_n^{+}(\omega)-f_n^{-}(\omega)+\frac{1}{2\pi (b^2)^n}~ 2 \pi i ~ \text{Res}\Big[e^{i \omega \theta} S^{(n)}(\theta)\Big]_{\theta=0}=0~~~.
\end{equation}

\begin{figure}[H]
\label{Contours}
	\centering
	\includegraphics[scale=1]{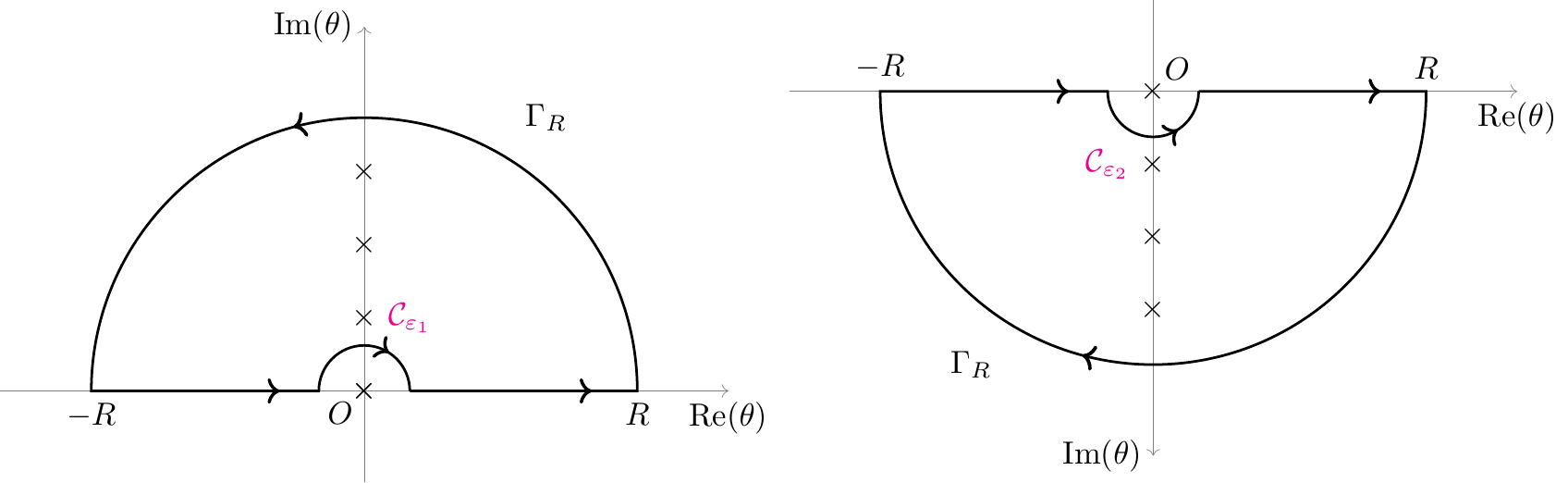}
\caption{Contours for perturbative retarded and advanced celestial amplitudes}	
\end{figure}

From the fig.\ref{Contours}, we see that the difference between $f_n^{+}(\omega)$ and $f_n^{-}(\omega)$ is same as the negative of $\frac{1}{2\pi (b^2)^n}~ 2 \pi i ~\text{Res}\Big[e^{i \omega \theta} S^{(n)}(\theta)\Big]_{\theta=0}$. This is because when we enclose the contour in the upper half-plane, the small semicircle $\textcolor{magenta}{\mathcal{C}_{\varepsilon_1}}$ is in clockwise sense which adds to the  $\textcolor{magenta}{\mathcal{C}_{\varepsilon_2}}$ semicircle in clockwise sense after taking the difference of $f_n^{+}(\omega)$ and $f_n^{-}(\omega)$ and by convention while calculating the residue at $\theta=0$ we enclose the semicircle in anticlockwise sense.    

As a consistency check we satisfy this equation by evaluating the residue of $e^{i \omega \theta} S^{(n)}(\theta)$ at $\theta=0$ for $n=1, \ldots, 5$. 
\begin{equation}
\small 
\begin{split}
\frac{1}{2\pi b^2}~ 2 \pi i ~ \text{Res}\Big[e^{i \omega \theta} S^{(1)}(\theta)\Big]_{\theta=0}&=\frac{1}{4}\\
\frac{1}{2\pi (b^2)^2}~ 2 \pi i ~ \text{Res}\Big[e^{i \omega \theta} S^{(2)}(\theta)\Big]_{\theta=0}&=\frac{\pi \omega-1}{32 \pi}\\
\frac{1}{2\pi (b^2)^3}~ 2 \pi i ~ \text{Res}\Big[e^{i \omega \theta} S^{(3)}(\theta)\Big]_{\theta=0}&=\frac{3\pi^2\omega^2-12\pi \omega+2\pi^2+6}{1536\pi^2}\\
\frac{1}{2\pi (b^2)^4}~ 2 \pi i ~ \text{Res}\Big[e^{i \omega \theta} S^{(4)}(\theta)\Big]_{\theta=0}&=\frac{\pi^3\omega^3-9\pi^2\omega^2+2\pi^3\omega+18\pi \omega-6\pi^2-6}{12288\pi^3}\\
\frac{1}{2\pi (b^2)^5}~ 2 \pi i ~ \text{Res}\Big[e^{i \omega \theta} S^{(5)}(\theta)\Big]_{\theta=0}&=\frac{5\pi^4 \omega^4-80\pi^3\omega^3+20\pi^4\omega^2+360\pi^2\omega^2-160\pi^3\omega-480\pi \omega+16\pi^4+240\pi^2+120}{1966080 \pi^4}~~~.
\end{split}
\end{equation}

Using crossing and unitarity conditions in eq.\eqref{boot1} and eq.\eqref{boot2} along with eq.\eqref{diff} we get 
\begin{equation}
\label{boot}
\begin{split}
&f_n^{+}(\omega)=\frac{1}{(1+e^{-\pi \omega})}\Bigg[-\underbrace{\frac{e^{-\pi \omega}}{2\pi (b^2)^n}~ 2 \pi i ~ \text{Res}\Big[e^{i \omega \theta} S^{(n)}(\theta)\Big]_{\theta=0}}_{\textcolor{blue}{\text{not fixed by crossing and unitarity conditions}}}-\underbrace{\sum_{j=1}^{n-1} \int_{-\infty}^{\infty}d\omega^{\prime}   f_{n-j}^{+}(\omega+\omega^{\prime}) f_j^{-}(\omega^{\prime})}_{\textcolor{blue}{\text{fixed by crossing and unitarity conditions}}}\Bigg]~~~.
\end{split}
\end{equation}

The second term in eq.\eqref{boot} is fixed by the crossing and unitarity conditions while the first term is not fixed by the crossing and unitarity conditions.

Therefore, we can calculate $f_2^{+}$, $f_3^{+}$ and so on
\begin{equation}
\small 
\begin{split}
&f_2^{+}(\omega)=\frac{1}{(1+e^{-\pi \omega})}\Bigg[-\frac{e^{-\pi \omega}}{2\pi (b^2)^n}~ 2 \pi i ~ \text{Res}\Big[e^{i \omega \theta} S^{(2)}(\theta)\Big]_{\theta=0}-\int_{-\infty}^{\infty}d\omega^{\prime}   f_{1}^{+}(\omega+\omega^{\prime}) f_1^{-}(\omega^{\prime})\Bigg]\\
&f_3^{+}(\omega)=\frac{1}{(1+e^{-\pi \omega})}\Bigg[-\frac{e^{-\pi \omega}}{2\pi (b^2)^n}~ 2 \pi i ~ \text{Res}\Big[e^{i \omega \theta} S^{(3)}(\theta)\Big]_{\theta=0}-\int_{-\infty}^{\infty}d\omega^{\prime}\Big[f_1^+(\omega+\omega^{\prime})f_2^-(\omega^{\prime})+f_2^+(\omega+\omega^{\prime})f_1^-(\omega^{\prime})\Big]\Bigg]\\
&f_4^{+}(\omega)=\frac{1}{(1+e^{-\pi \omega})}\Bigg[-\frac{e^{-\pi \omega}}{2\pi (b^2)^n}~ 2 \pi i ~ \text{Res}\Big[e^{i \omega \theta} S^{(4)}(\theta)\Big]_{\theta=0}\\
&~~~~~~~~~~~~~~~~~~~~~~~~~~~-\int_{-\infty}^{\infty}d\omega^{\prime}\Big[f_1^+(\omega+\omega^{\prime})f_3^-(\omega^{\prime})+f_2^+(\omega+\omega^{\prime})f_2^-(\omega^{\prime})+f_3^+(\omega+\omega^{\prime})f_1^-(\omega^{\prime})\Big]\Bigg]\\
&f_5^{+}(\omega)=\frac{1}{(1+e^{-\pi \omega})}\Bigg[-\frac{e^{-\pi \omega}}{2\pi (b^2)^n}~ 2 \pi i ~ \text{Res}\Big[e^{i \omega \theta} S^{(4)}(\theta)\Big]_{\theta=0}\\
&~~~~~~~~~~~~~~~~~-\int_{-\infty}^{\infty}d\omega^{\prime}\Big[f_1^+(\omega+\omega^{\prime})f_4^-(\omega^{\prime})+f_2^+(\omega+\omega^{\prime})f_3^-(\omega^{\prime})+f_3^+(\omega+\omega^{\prime})f_2^-(\omega^{\prime})+f_4^+(\omega+\omega^{\prime})f_1^-(\omega^{\prime})\Big]\Bigg]~~~.
\end{split}
\end{equation}

\section{Gravitational dressing of the $2d$ QFT amplitude in celestial space}
\label{gravdres}
For integrable field theories, in presence of irrelevant deformation $T\bar{T}$, the $2d$ QFT $\mathcal{S}$-matrix is modified by a pure phase \cite{Cavaglia:2016oda} 

\begin{equation}
S_{ij}^{kl}(\theta) \to S_{ij}^{kl}(\theta) e^{i\delta_{ij}^{(t)}(\theta)}~~~, 
\end{equation}

where, the diagonal phase shift $\delta_{ij}^{(t)}(\theta)$ is given by deformation parameter $t$ and difference of rapidities $\theta=\theta_i-\theta_j$
\begin{equation}
\delta_{ij}^{(t)}(\theta)=t m_i m_j \sinh \theta~~~.
\end{equation}
The deformation parameter $t$ is related to the string length $t=2l_s^2$ in effective string theory. For same mass particles, we have
\begin{equation}
S(\theta) \to S(\theta) e^{2i l_s^2 m^2 \sinh\theta}~~~.
\end{equation}  

This factor is also discussed in the paper \cite{Paulos:2016but} as a solution to the crossing and unitarity conditions i.e., $S(\theta)=S(i\pi-\theta)$ and $S(\theta)S(-\theta)=1$~.

In terms of $s$-variable we have

\begin{equation}
S(s) \to S(s) e^{il_s^2 \sqrt{s(s-4m^2)}}~~~,
\end{equation}
the dressing factor for massless particle reduces to $e^{i l_s^2s}$.

This dressing factor introduced in \cite{Dubovsky:2013ira} in the context of gravitational scattering of relativistic point particles in trans-Planckian regime and large impact parameter is referred to as the gravitational dressing factor.
 
In this section, we study the gravitational dressing of the $2d$ QFT amplitude in the celestial space restricting to massless particles. 

The gravitational dressing of the $2d$ QFT amplitude for massless particles is given by
\begin{equation}
\begin{split}
S(s) \to S(s) S_{\text{grav}}(s)~~~,
\end{split}
\end{equation}
where, the gravitational dressing factor is 
\begin{equation}
S_{\text{grav}}(s)=e^{i l_s^2s}~~~.
\end{equation}

The $\mathcal{S}$-matrix is a function of mandelstem variable $s$ which is center-of-mass energy squared.  
In the celestial space we define the celestial amplitude $\mathcal{A}(\omega)$ as  
\begin{equation}
\begin{split}
\mathcal{A}(\omega)\equiv& \int_{0}^{\infty} ds s^{ \omega-1} S(s)~~~\\
\implies S(s)&=\frac{1}{2 \pi i} \int_{\gamma-i\infty}^{\gamma+i\infty}d\omega s^{-\omega}\mathcal{A}(\omega)~~~,
\end{split}
\end{equation}
where, we trade the mandelstem variable $s$ for a rindler energy or conformal dimension $\omega$ diagonalizing boosts in the directions of the particles.

Now, Mellin transform of $e^{i l_s^2 s}$ is 
\begin{equation}
\begin{split}
\int_{0}^{\infty} ds s^{ \omega-1} e^{il_s^2s}=(-i l_s^2)^{-\omega} \Gamma(\omega)~~~.
\end{split}
\end{equation}
The gravitational dressing gives
\begin{equation}
\begin{split}
\int_{0}^{\infty} ds s^{ \omega-1} ~S(s) \to \int_{0}^{\infty} ds s^{ \omega-1} ~S(s) e^{i l_s^2 s}
\end{split}
\end{equation}

	The gravitational dressing condition becomes
	\begin{equation}
	\begin{split}
	\mathcal{A}(\omega)
	& \to \frac{1}{2\pi i} \int_{\gamma-i\infty}^{\gamma+i\infty} d\omega^{\prime} (-i l_s^2)^{-\omega^{\prime}} \Gamma(\omega^{\prime})\mathcal{A}(\omega-\omega^{\prime})~~~.\\
	\end{split}
	\end{equation}


Now, we take several ansatzes for $\mathcal{A}(\omega)$ that has a pole on the right half-plane of $\omega$ and see what we get after the convolution.
\subsection*{Ansatz for $\mathcal{A}(\omega)$}
\subsection*{Ansatz 1}

We take $\mathcal{A}(\omega)$ as 

\begin{equation}
\mathcal{A}(\omega)=\csc\pi \omega~~~.
\end{equation}
The function has poles at $\omega=n~,~~n\in \mathbb{Z}~~.$ 

Performing the convolution we get 
\begin{equation}
\label{convolution1}
\begin{split}
&\frac{1}{2\pi i} \int_{\gamma-i\infty}^{\gamma+i\infty} d\omega^{\prime} (-i l_s^2)^{-\omega^{\prime}} \Gamma(\omega^{\prime})\csc(\pi(\omega-\omega^{\prime}))\\
&=\frac{e^{-il_s^2} \Gamma(\omega)\Gamma(1-\omega,-il_s^2)}{\pi}~~~.
\end{split}
\end{equation}
where we use the Mellin-Barnes integral representation of the upper incomplete gamma function $\Gamma(a,z)$[\textbf{Eq.(3.4.11)} in \cite{Paris}, p. 113] 
\begin{equation}
\label{Mellin-Barnes1}
\Gamma(a,z)=-\frac{z^{a-1}e^{-z}}{\Gamma(1-a)} \frac{1}{2 \pi i} \int_{\gamma-i\infty}^{\gamma+i\infty} ds ~\Gamma(s+1-a) \pi z^{-s} \csc \pi s~~~.
\end{equation}
we put $s=\omega^{\prime}-\omega,~a=1-\omega,~z=-i l_s^2$ in eq.\eqref{Mellin-Barnes1} to prove eq.\eqref{convolution1}.

Now, the upper incomplete gamma function $\Gamma(a, z)$ is an entire function of $a$ when $z \neq 0$. 
Therefore, for $l_s^2 \neq 0$, $\Gamma(1-\omega, -il_s^2)$ is an entire function of $1-\omega$. The function $\Gamma(\omega)$ has poles at $\omega=n~,~~n=\mathbb{Z}^{-} \displaystyle \cup \{0\}$.

$\csc \pi \omega$ has poles at $\omega=n~,~~n\in \mathbb{Z}$, after gravitational dressing the function $\frac{e^{-il_s^2} \Gamma(\omega)\Gamma(1-\omega,-il_s^2)}{\pi}$ has poles at $\omega=n~,~~n=\mathbb{Z}^{-} \displaystyle \cup \{0\}$.

We can see that after gravitational dressing the poles on the right half-plane of $\omega$ are absent.	
\subsection*{Ansatz 2}
We take $\mathcal{A}(\omega)$ as
\begin{equation}
\mathcal{A}(\omega)=\frac{1}{\omega-\mathfrak{C}}~~~,
\end{equation}
where, $\mathfrak{C}$ is located on the right half plane, i.e., $\text{Re}(\mathfrak{C})>0$.

Performing the convolution we get
\begin{equation}
\begin{split}
&\frac{1}{2\pi i} \int_{\gamma-i\infty}^{\gamma+i\infty} d\omega^{\prime} (-i l_s^2)^{-\omega^{\prime}} \Gamma(\omega^{\prime})\frac{1}{\omega-\omega^{\prime}-\mathfrak{C}}~~~.\\
&=(-il_s^2)^{\mathfrak{C}-\omega} \gamma(\omega-\mathfrak{C},-il_s^2)~~~.
\end{split}
\end{equation}

where we use the Mellin-Barnes integral representation of the  lower incomplete gamma function $\gamma(a,z)$ [\textbf{Eq.(3.4.10)} in \cite{Paris}, p. 113]

\begin{equation}
\begin{split}
\gamma(a,z)=\frac{1}{2\pi i} \int_{\gamma-i\infty}^{\gamma+i\infty} ~ds ~\frac{\Gamma(-s)}{s+a} z^{s+a} ds ~~~.
\end{split}
\end{equation}

Now, the  lower incomplete gamma function $\gamma(\omega-\mathfrak{C},-il_s^2)$ is meromorphic with simple poles at 
\begin{equation}
\begin{split}
\omega-\mathfrak{C}&=-n~~~,~~~n=\mathbb{Z}^{+} \displaystyle \cup \{0\}\\
\implies \omega &=\mathfrak{C}-n~~~.
\end{split}
\end{equation} 

For $n=0$~, ~$\text{Re}(\mathfrak{C})>0$ therefore we have a pole at $\omega=\mathfrak{C}$ which lies on the right half-plane.

Now, if we impose $\text{Re}(\mathfrak{C}-n)<0$~,~ $n=\mathbb{Z}^{+}$ then we say that there are no other poles on the right half-plane but still there is a pole at $\omega=\mathfrak{C}$, $\text{Re}(\mathfrak{C})>0$.

\subsection*{Ansatz 3}
We take $\mathcal{A}(\omega)$ as 

\begin{equation}
\mathcal{A}(\omega)=\Gamma(-\omega)~~~.
\end{equation}
The function has poles at $\omega=n~,~~n=\mathbb{Z}^{+} \displaystyle \cup \{0\}~~~.$


Performing the convolution we get 
\begin{equation}
\begin{split}
&\frac{1}{2\pi i} \int_{\gamma-i\infty}^{\gamma+i\infty} d\omega^{\prime} (-i l_s^2)^{-\omega^{\prime}} \Gamma(\omega^{\prime})\Gamma(-(\omega-\omega^{\prime}))\\
&=2 \left(-i l_s^2\right)^{-\frac{\omega }{2}} K_{\omega }\left(2 \sqrt{-i l_s^2}\right)~~~.
\end{split}
\end{equation}
where we use the Mellin-Barnes integral representation of the modified Bessel function of the second kind $K_{\nu}(z)$[\textbf{Eq.(3.4.18)} in \cite{Paris}, p. 114] 
\begin{equation}
K_{\nu}(z)=\frac{\Big(\frac{1}{2}z\Big)^{\nu}}{4 \pi i} \int_{\gamma-i\infty}^{\gamma+i\infty} ds ~\Gamma(s)\Gamma(s-\nu) \Big(\frac{z}{2}\Big)^{-2t}  ~~~.
\end{equation}
Now, the modified Bessel function of the second kind $K_{\nu}(z)$ has only one singular point at $\nu =\infty$ for fixed $z$. Here, for fixed $l_s^2$, $K_{\omega } (2 \sqrt{-i l_s^2})$ has pole when $\omega \to \infty$.

We can see that after gravitational dressing the poles on the right half-plane of $\omega$ are absent.

\par 
Therefore, from this analysis we see that the poles on the right half-plane are absent after gravitational dressing if the function contains multiple poles.  

\par 
The absence of poles has nice analogy with the absence of the bulk point singularity for AdS/CFT correlators \cite{Maldacena:2015iua}. For celestial amplitude, in \cite{Arkani-Hamed:2020gyp} we see that the exponentially soft high-energy behavior when translated into celestial amplitude erases the poles on the right half-plane. We can think the gravitaional dressing condition in $2d$ as a constraint which gives some constraint on the analytic structure of the celestial amplitude. 
\section{Conclusions and furure directions}
\label{conc}
In this paper, we study celestial amplitude for $2d$ bulk $\mathcal{S}$-matrix and show that for massive scalar particles the celestial amplitude is just the fourier transform of the $\mathcal{S}$-matrix written in the rapidity variable. For the Sinh-Gordon $\mathcal{S}$-matrix we evaluate the perturbative celestial amplitude and see that there should be two types of celestial amplitude, the retarded and the advanced due to the presence of the pole in the origin of the complex rapidity-plane. 
\par 
Here, for the Sinh-Gordon model, the exact $\mathcal{S}$-matrix has no pole at rapidity, $\theta=0$, since $\alpha$ is real. The rapidity, $\theta=0$ pole is a perturbative manifestation, for this there are these two perturbative celestial amplitudes corresponding to
two $i\epsilon$ prescriptions. The difference between these two perturbative amplitudes is related to the residue of the pole at $\theta=0$. One important point to note is that these two perturbative amplitudes also naturally appear in unitarity equation. After resumming the perturbative celestial amplitude it's nice to get the celestial amplitude corresponding to the exact $\mathcal{S}$-matrix. Since there are two perturbative amplitudes, which one will resum to be the celestial amplitude for the exact $\mathcal{S}$-matrix? This is a very interesting question. To evaluate the celestial amplitude corresponding to the exact $\mathcal{S}$-matrix by resummation of the perturbative celestial amplitude is difficult. 
\par 
In the celestial space, we translate the crossing and unitarity conditions and check these conditions for the Sinh-Gordon model. From the celestial CFT perspective, we ask about determining the higher order celestial amplitude from the lower order i.e.,  bootstrapping celestial amplitude  by imposing the ``crossing and unitarity'' conditions. Finally, we analyze the gravitational dressing condition for the $2d$ QFT amplitude in celestial space restricted to massless particles. We see that this condition manifests itself as eraser of poles from the right half-plane in the celestial space. There are several promising future directions related to this work.

\par
\textbf{Bootstrapping celestial amplitude in $4d$ for massive scalar theory:} It would be interesting to extend our analysis to the massive $\phi^4$ theory in $4d$ with an aim to compute the one-loop correction to the four-point amplitude in $\phi^4$ theory in flat space by imposing the ``crossing and unitarity conditions'' to the celestial amplitude. 

\par 
\textbf{Gravitational dressing for massive scaterring in $2d$:}
Another question is to translate the gravitational dressing for massive scattering in $2d$ into celestial space and see it's consequence. In doing this, the exact analytical result for inverting the gravitational dressing factor for the massive case in the celestial space can't be obtained. But, one can try to analyze this using some approximation.

\par

\textbf{Space of celestial amplitude in 2d:} Another interesting direction in massive QFTs in $2d$ is to find the allowed dual celestial space of $2 \to 2$ $\mathcal{S}$-matrix by imposing the crossing, unitarity, analyticity and $O(N)$ symmetry into the celestial amplitude i.e., to understand the dual space of \cite{Cordova:2019lot} in the celestial space.   

\par
\textbf{Flat limit of massive scalar AdS amplitude and it's contact with massive scalar celestial amplitude:} 
For the massive scalar celestial amplitude, it's nice to connect the celestial amplitude to the flat limit of AdS amplitude. For this, we can use the dictionary between the positions of operators at the boundary of AdS and the momenta of particles in the flat limit of AdS \cite{Komatsu:2020sag}. The celestial amplitude written in conformal basis which is an integral transform of the flat space amplitude in momentum space \cite{Pasterski:2017kqt} can be translated to an integral transform of positions of operators at the boundary of AdS. We can then try to see the interpretation of the integral transform in terms of CFT living on the boundary. The light-ray operators are some class of operators in CFT defined by an integral, it would be interesting to make connections with the integral transform with light-ray operators. In this context, we can try to perform the integral transform by taking some AdS contact correlators and analyze the result.  
\par
\textbf{Flat limit of the massless scattering in AdS:}
We can take the bulk-point limit related to high energy scattering of the massive scalar celestial amplitude expressed in the boundary of AdS in which the mass of the particles can be neglected. It's nice to make a connection with the flat limit of massless scattering in AdS \cite{Okuda:2010ym}. 
  
\par
\textbf{Conformal block expansion for the celestial amplitude:} 
In this line of thinking, it would be useful to understand the analog of conformal block expansion for the celestial amplitude. In AdS, we can convolute the conformal block expansion of the boundary four-point function with the integral transform and see how the block gets modified. We can work entirely in flat space and using the partial wave expansion of the four point amplitude in a basis of Gegenbauer polynomials \cite{Chaichian} we can perform the integral transform and the question is after the integral transform whether the transformed object has a nice structure in terms of conformal block decomposition, it would be nice to make connection with the conformal block expansion in the Celestial CFT \cite{Atanasov:2021cje}.      
\section*{Acknowledgements}
I would like to thank Shota Komatsu for his guidance throughout this work and comments on the earlier version of the draft. I also thank R. Loganayagam for useful discussions, giving me mental support and comments on the earlier version of the draft. I also thank Chethan Krishnan and Suvrat Raju for useful discussions. This work is supported by the Department of Atomic Energy, Government of India, under project no. RTI4001. 
\appendix
\section{Perturbative retarded and advanced celestial amplitude computation in $2d$ Sinh-Gordon model}
\label{pert}

In this appendix \ref{pert}, we calculate the perturbative retarded and advanced celestial amplitude in $2d$ Sinh-Gordon model. 

In the complex $\theta$-plane, the perturbative $\mathcal{S}$-matrices $S^{(n)}(\theta)$ for $n=1, \ldots, 5$ contain poles at $\theta=n\pi i$, $n \in \mathbb{Z}$. 

\begin{figure}[H]
	\centering
	\includegraphics[scale=1]{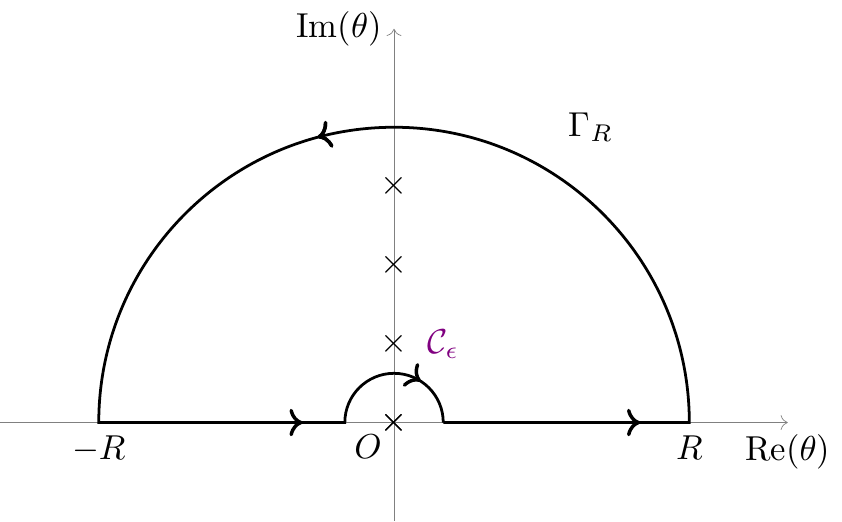}
	\caption{Contour for evaluating perturbative retarded celestial amplitude}
	\label{con1}	
\end{figure}

While calculating the perturbative retarded celestial amplitude $f_n^{+}(\omega)$, we choose the contour in the upper half-plane as in fig.\ref{con1} and enclose the poles as $\theta=n \pi i$, $n \in \mathbb{Z}^{+}$.

\begin{figure}[H]
	\centering
	\includegraphics[scale=1]{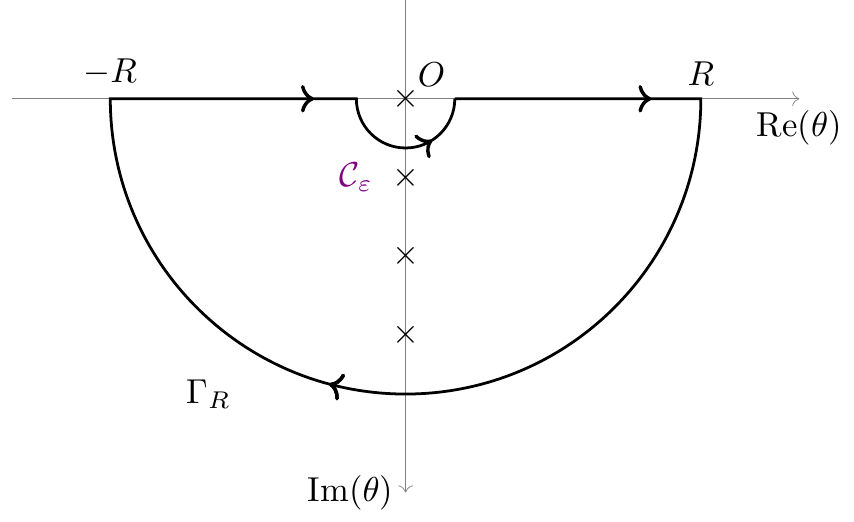}
\caption{Contour for evaluating perturbative advanced celestial amplitude}
\label{con2}	
\end{figure}

While calculating the perturbative advanced celestial amplitude $f_n^{-}(\omega)$, we choose the contour in the lower half-plane as in fig.\ref{con2} and enclose the poles as $\theta=n \pi i$, $n \in \mathbb{Z}^{-}$.

The residues of the functions appeared in the perturbative expansion of the $\mathcal{S}$-matrix at the poles as $\theta=n \pi i$, $n \in \mathbb{Z}$ are given by 
\begin{equation}
\begin{split}
\text{Res}\Big[\csch\theta~ e^{i \omega \theta} \Big]_{\theta=n\pi i}&=(-1)^n e^{-n\pi \omega}\\
\text{Res}\Big[\csch^2\theta~ e^{i \omega \theta} \Big]_{\theta=n\pi i}&=i\omega~ e^{-n\pi \omega}\\
\text{Res}\Big[\csch^3\theta~ e^{i \omega \theta} \Big]_{\theta=n\pi i}&=-\frac{1}{2}(-1)^n(\omega^2+1)e^{-n\pi \omega}\\
\text{Res}\Big[\csch^4\theta~ e^{i \omega \theta} \Big]_{\theta=n\pi i}&=-\frac{1}{6}i\omega (\omega^2+4)e^{-n\pi \omega}\\
\text{Res}\Big[\csch^5\theta~ e^{i \omega \theta} \Big]_{\theta=n\pi i}&=\frac{1}{24}(-1)^n(\omega^2+1)(\omega^2+9)e^{-n\pi \omega}~~~.
\end{split}
\end{equation}






\newpage 
\providecommand{\href}[2]{#2}\begingroup\raggedright
\endgroup

\end{document}